\documentclass[aps,twocolumn,floatfix,superscriptaddress,prl]{revtex4-1}
\usepackage{amssymb}
\usepackage{graphicx}
\usepackage{dcolumn}
\usepackage{amsmath}
\usepackage{bm}
\usepackage{yhmath}
\usepackage{textcomp}
\usepackage{epstopdf}
\usepackage{color}
\usepackage{ulem}
\usepackage[toc,page,title,titletoc,header]{appendix}
\usepackage[colorlinks, 
            linkcolor=blue,
            anchorcolor=blue,
            citecolor=blue
            ]{hyperref}
\usepackage{natbib}
\newcommand{\bfe}{{\bf e}}
\newcommand{\bfr}{{\bf r}}
\newcommand{\bfk}{{\bf k}}

\newcommand{\bfq}{{\bf q}}

\newcommand{\bfQ}{{\bf Q}}

\newcommand{\bfn}{{\bf n}}
\newcommand{\bfm}{{\bf m}}
\newcommand{\bfK}{{\bf K}}
\newcommand{\hbfx}{\hat{\bf x}}
\newcommand{\hbfy}{\hat{\bf y}}
\newcommand{\hbfz}{\hat{\bf z}}
\newcommand{\tlc}{\tilde{c}}

\newcommand{\be}{\begin{eqnarray}}
\newcommand{\ee}{\end{eqnarray}}

\begin{document}

\title{Emergence of Metallic Quantum Solid Phase in a Rydberg-Dressed Fermi Gases}

\author{Wei-Han Li}
\affiliation{Department of Physics, National Tsing-Hua University, Hsinchu, Taiwan}
\author{Tzu-Chi Hsieh}
\affiliation{Department of Physics, National Tsing-Hua University, Hsinchu, Taiwan}
\author{ Chung-Yu Mou}
\affiliation{Department of Physics, National Tsing-Hua University, Hsinchu, Taiwan}
\affiliation{Physics Division, National Center for Theoretical Sciences, Hsinchu,Taiwan}
\author{Daw-Wei Wang}
\affiliation{Department of Physics, National Tsing-Hua University, Hsinchu, Taiwan}
\affiliation{Physics Division, National Center for Theoretical Sciences, Hsinchu,Taiwan}

\begin{abstract}
We examine possible low-temperature phases of a repulsively Rydberg-dressed Fermi gas in a three-dimensional free space. It is shown that the collective density excitations develop a roton minimum, which is softened at a wavevector smaller than the Fermi wavevector when the particle density is above a critical value. The mean field calculation shows that unlike the insulating charge density waves states often observed in conventional condensed matters, a self-assembled metallic density wave state emerges at low temperatures. In particular, the density wave state supports a Fermi surface and a body-center-cubic crystal order at the same time with the estimated critical temperature being about one-tenth of the non-interacting Fermi energy. Our results suggest the emergency of a fermionic quantum solid that should be observable in current experimental setup.
\end{abstract}

\maketitle
{\it Introduction:} 
It is well known that the system of repulsively interacting Fermi gases is mainly controlled by the celebrated Fermi liquid (FL) theory\cite{Landau1}. The breakdown of the FL theory can lead to exotic self-organizing orders even without the presence of lattice potentials. For example, in the strong interaction regime, the ground state may become unstable to a nematic state by breaking the rotational symmetry via a Fermi surface distortion (Pomeranchuk instability, PI \cite{PI}). For systems of a long-ranged Coulomb/dipolar interaction, it is known that particles can be "frozen" locally without exchanging their positions and form a classical crystal with one particle per site in the dilute/dense limit \cite{Wigner,Gruner,Baranov}. In another extreme situations, such as high density $^3He$ under pressure, the ground state can be even turned from the Fermi liquid into a quantum solid \cite{solidhe3,quantum_solid}, where particles self-assemble crystal order but are still intrinsically restless and exchanging their positions even at the absolute zero of temperature.

\begin{figure}[tbp]
\includegraphics[width=8cm]{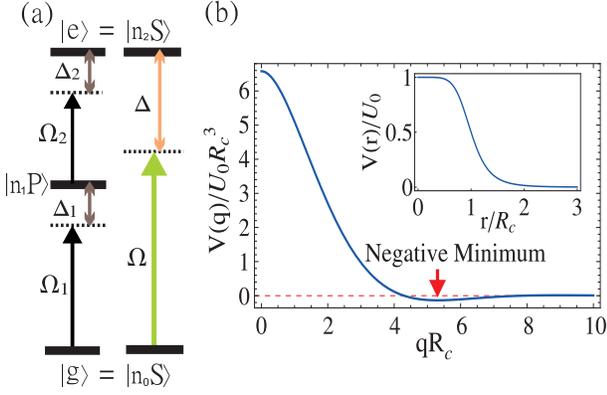} \centering
\caption{
(a) Schematic plot for a two-photon excitation for an off-resonant coupling between the ground state ($|n_{0}S\rangle$) and a Rydberg state ($|n_{2}S\rangle$) via an intermediate state ($|n_{1}P\rangle$). $\Delta_{1,2}$ and $\Omega_{1,2}$ are the detuning and Rabi frequency respectively. An effective single photon expression can be  obtained with $\Delta=\Delta_{1}+\Delta_{2}$ and $\Omega=\Omega_{1}\Omega_{2}/2\Delta_{1}$.
(b) Effective Rydberg-dressed interaction in the far detuning limit ($\Delta\gg\Omega$) in the momentum space. The insert shows the corresponding real space profile with the blockade radius, $R_{C}$.}
\label{Rydberg_energy_level}
\end{figure}

However, in the traditional condensed matter systems, these interesting phases (nematic state, classical crystal, or quantum solid) cannot be achievable easily because the interaction strength has to be strong enough to compete with the Fermi energy. On the other hand, in the system of Rydberg atoms, the length scale and strength of the effective inter-atom interaction can be manipulated easily by external fields \cite{Dudin,Raitzsch,Pritchard,Nipper,Schaus}. In addition to the blockade effect for on-resonant excitations \cite{Jaksch,Lukin,Urban,Gaetan}, one can also apply a far-detuned weak field (see Fig\ref{Rydberg_energy_level} (a)) to generate an effective Rydberg-dressed interaction (RDI), which replaces the short-ranged Lennard-Jones potential by a soft core with finite interaction range (see in Fig\ref{Rydberg_energy_level}(b)) \cite{Henkel1,Henkel2,Balewski}. Theoretical calculations show that a repulsive RDI in a Bose gas may lead to a supersolid droplet phase \cite{boson_crystal,Cinti,Pupillo,Henkel1,Henkel2}, while an attractive RDI induces a 3D bright soliton \cite{Maucher}. For a Rydberg Fermi gas, some topological phases are also predicted for an attractive \cite{Xiong} or repulsive interaction in a optical lattice near half-filling \cite{Li}.

In this paper, we demonstrate that a self-assembled metallic quantum solid phase can emerge in a single species Fermi gas even for a {\it weakly} repulsive RDI in a 3D continuous space. The quantum phase transition from Fermi liquid to quantum solid is mainly driven by the interaction range of the RDI, and is shown to be a first order transition near the collective mode softening point at a wavevector smaller than the Fermi wavevector. The new ground state, mettalic quantum solid, has a gapless fermionic excitation on top of a density wave order, which has a body-centered-cubic (BCC) structure with a lattice constant a few times larger than the averaged inter-particle distance, i.e. each unit cell contains many and non-integer fermionic atoms to form a Fermi sea. We further estimate the critical temperature of the density wave order about 0.1$E_F^0$, where $E_F^0$ is the noninteracting Fermi energy. Our results indicate a new quantum order originating solely from the finite interaction range, and open a new pathway to form non-conventional correlated quantum states.

{\it Effective Interaction:}
In this paper, we consider a single-species Fermi gas, taking $^{6}Li$ as an example, where each atom is weakly coupled to its $s$-wave Rydberg state by an off-resonant two photon transition via an intermediate state, see Fig.\ref{Rydberg_energy_level}(a). In the far detuning and weak coupling limit, we can apply the standard perturbative and adiabatic approximation\cite{Henkel1} to obtain the effective interaction between dressed state atoms through the effective Raman coupling $\Omega$ and detuning $\Delta$: the Rydberg-dressed interaction (RDI) can be expressed to be, $V_{\rm RD}(\mathbf{r})=\frac{U_{0}}{1+(r/R_{c})^{6}}$\cite{Henkel2,Maucher,Honer,Xiong}, where $U_{0}\equiv(\Omega/2\Delta)^{4} C_{6}/R_{c}^{6}$, and $R_{c}\equiv (C_{6}/2|\Delta|)^{1/6}$ are the interaction strength and the averaged soft-core radius respectively. $C_{6}$ is the averaged van der Waals coefficient, which can be shown to be positive for all orbital states when exciting $^{6}Li$ to a state with $n>30$ \cite{Walker,Singer2}. 

We note that the effective Rydberg-dressed interaction discussed above is more justified for the single-species Fermi gas \cite{Xiong} than for bosonic systems, because the Pauli exclusion principle can strongly reduce the possible atomic loss due to the orbital level crossing in the short-distance regime. As a result, we can calculate the scattering amplitude in 3D free space (valid in the weak interaction limit, $U_{0}/E_{F}^0\ll k_{F}^0R_{c}$ \cite{Landau2}) by the first Born approximation: $V(\mathbf{q})\equiv\int d\bfr\,V_{\rm RD}(\bfr)\,e^{-i\bfq\cdot\bfr}=U_{0}R_{c}^{3}\tilde{V}(|\bfq|R_c)$, where
\begin{equation}
\tilde{V}(s)=\frac{2\pi^{2}}{3s}\left[1+2\,e^{s/2}\sin\left(\frac{\sqrt{3}s}{2}-\frac{\pi}{6}\right)\right]\,e^{-s}.
\label{Vq}
\end{equation}
is a single-parameter function with $s\equiv |\bfq|R_c$. As shown in Fig.\ref{Rydberg_energy_level}(b), such scattering amplitude has a negative minimum at a finite wavevector, $|\bfq|=Q_{c}\sim 5.3/R_c$. This spacial property results from the blockade effects of the Rydberg-dressed interaction in real space (see the inset), and does not exist in other kinds of long ranged interaction, say Coulomb or dipolar interaction. As we will show below, it helps to stabilize the density wave state even for Fermi gases.

{\it Collective density mode softening:}
To investigate the possible density wave order, we first calculate the retarded density correlation function, $D^{R}(\mathbf{q},\omega)\equiv -i\int_0^\infty dt e^{-i\omega t} \langle[\rho(\mathbf{q},t),\rho(\mathbf{-q},0)]\rangle$, with the density operator $\rho(\mathbf{q})\equiv\sum_\bfk c^{\dag}_{\mathbf{k+q}}c_{\mathbf{k}}$ ($c_\bfk$ is the field operator of Rydberg-dressed fermions). 
The density correlation function is directly related to the full polarizability, $\Pi(\mathbf{q},\omega)$ (${\rm Re} D^{R}={\rm Re}\Pi$, ${\rm Im} D^{R}={\rm sgn}(\omega){\rm Im}\Pi$, \cite{Fetter}), which can be evaluated through Dyson series:
\begin{equation}\label{PI}
\Pi(\mathbf{q},\omega)=\frac{\Pi^\ast(\mathbf{q},\omega)}{1-V(\bfq)\Pi^\ast(\mathbf{q},\omega)}
\end{equation}
with $\Pi^\ast(\bfq,\omega)$ being the irreducible polarizability.

In our present Rydberg-dressed system, we are interesting in the regime of long blockade radius (i.e. high density), i.e. $k_F^0R_c\gg 1$, so that the direct term contributes much larger than the exchange and correlation terms (similar to the Coulomb interaction case in the high density limit, see Ref. \cite{Fetter}). As a result, we can apply the random phase approximation (RPA) to replace the irreducible polariability ($\Pi^\ast$) by the noninteracting polarizability, $\Pi_{0}(\bfq,\omega)=\frac{-i}{(2\pi)^{4}}\int\,d^{3}k d\nu G_{0}(\bfk,\nu)G_{0}(\bf{k}+\bf{q},\omega+\nu)$, where $G_{0}(\bfk,\omega)$ is the noninteracting Green's function and can be evaluated analytically \cite{Fetter}. 

\begin{figure}
\includegraphics[width=9cm]{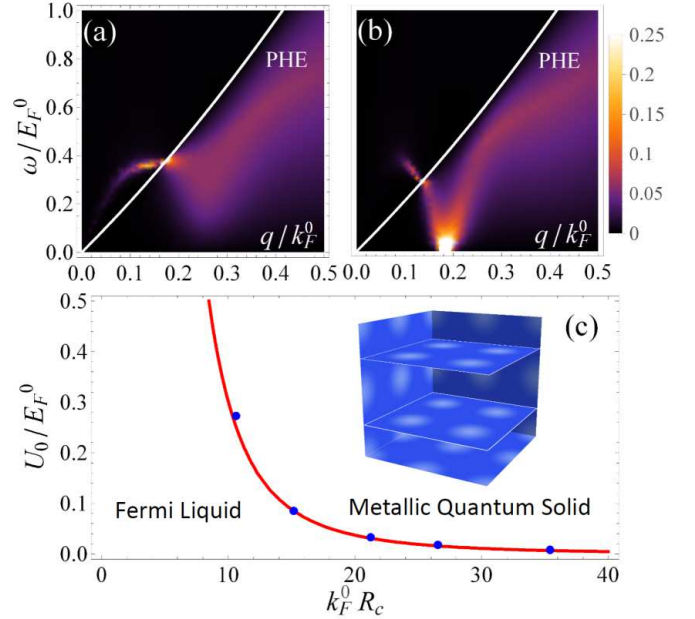} \centering
\caption{Spectral weights ($\propto  \Im\Pi(q,\omega)$) of the collective excitations for $U_{0}/E_{F}^0=0.012$ for (a)  $k_{F}^0R_{c}=20$ and (b) $k_{F}^0R_{c}=29$, respectively. The while thin lines indicate the regime of particle-hole excitations, and the roton minimum reaches zero energy in (b). (c) Phase diagram between the Fermi liquid and the metallic density wave. The red solid line is determined by the condition of collective mode softening, while the blue dots are obtained by numerically minimizing the total meanfield energy at zero temperature. The inset shows a schematic density plot of the metallic density wave with a BCC structure.}
\label{collective_mode}
\end{figure}
In Fig. \ref{collective_mode}(a), we show a typical spectral weight of the collective mode excitations within RPA. The dispersion of the collective mode is determined from the pole of $\Pi(\bfq,\omega)$ (see Eq. (\ref{PI})): $1=V(q)\Re\Pi_{0}(q,\omega_{q})$ for $\Im\Pi_0(q,\omega_{q})=0$ (i.e. outside the regime of particle-hole excitations (PHE) \cite{Mahan,Wang}). In the long wavelength limit, this mode has to a linear dispersion, $\omega_q=cq+{\cal O}(q^2)$, with the zero sound velocity, $c\equiv v_F^0\left[1+2e^{-2}\exp(-3/mk_F^0U_0R_c^{3})\right]$ ($v_F^0$ is the noninteracting Fermi velocity). In the shorter wavelength (or larger wavevector) regime, the collective excitation is strongly damped and broadened by particle-hole excitations.

However, when the interaction range ($R_c$) is tuned larger (or in higher density regime), we find that the roton excitation is softened at a finite momentum ($q=Q_c\sim 0.2 k_F^0$ as shown in Fig. \ref{collective_mode}(b)). In fact, since $\Im\Pi_0(q,\omega\to 0)\to 0$, the condition to have such roton softening can be analytically derived as $k_FR_c\gg 1$: 
\begin{equation}
\label{long length}
\frac{U_{0}}{E_{F}^0}\times(k_FR_{c})^{3}\geq\frac{4\pi ^{2}}{\mid \tilde{V}(5.3)\mid}\approx 294.6.
\end{equation}
In Fig. \ref{collective_mode}(c), we show the obtained quantum phase diagram, where the Landau's FL theory fails in the high density regime ($R_ck_F^0\gg 1$) even the interaction strength ($U_0/E_F^0$) is small. Note that this result is within the same condition to justify the first Born approximation and RPA used in our theoretical calculation. Similar results can be also obtained in lower dimensional systems, while the above two approximations cannot be justified. We emphasize that such interesting result does not appear in other long-ranged interaction (such as Coulomb or dipolar interaction), because the negative minimum of $V(\bfq)$ is originated from the sharp changes of the interaction profile due to blockade effects (see Fig. \ref{Rydberg_energy_level}(b)). We also have examed that there is no possibility to have PI in such Rydberg-dressed interaction within the parameter regime here.

\begin{figure}[tbp]
\includegraphics[width=7.5cm]{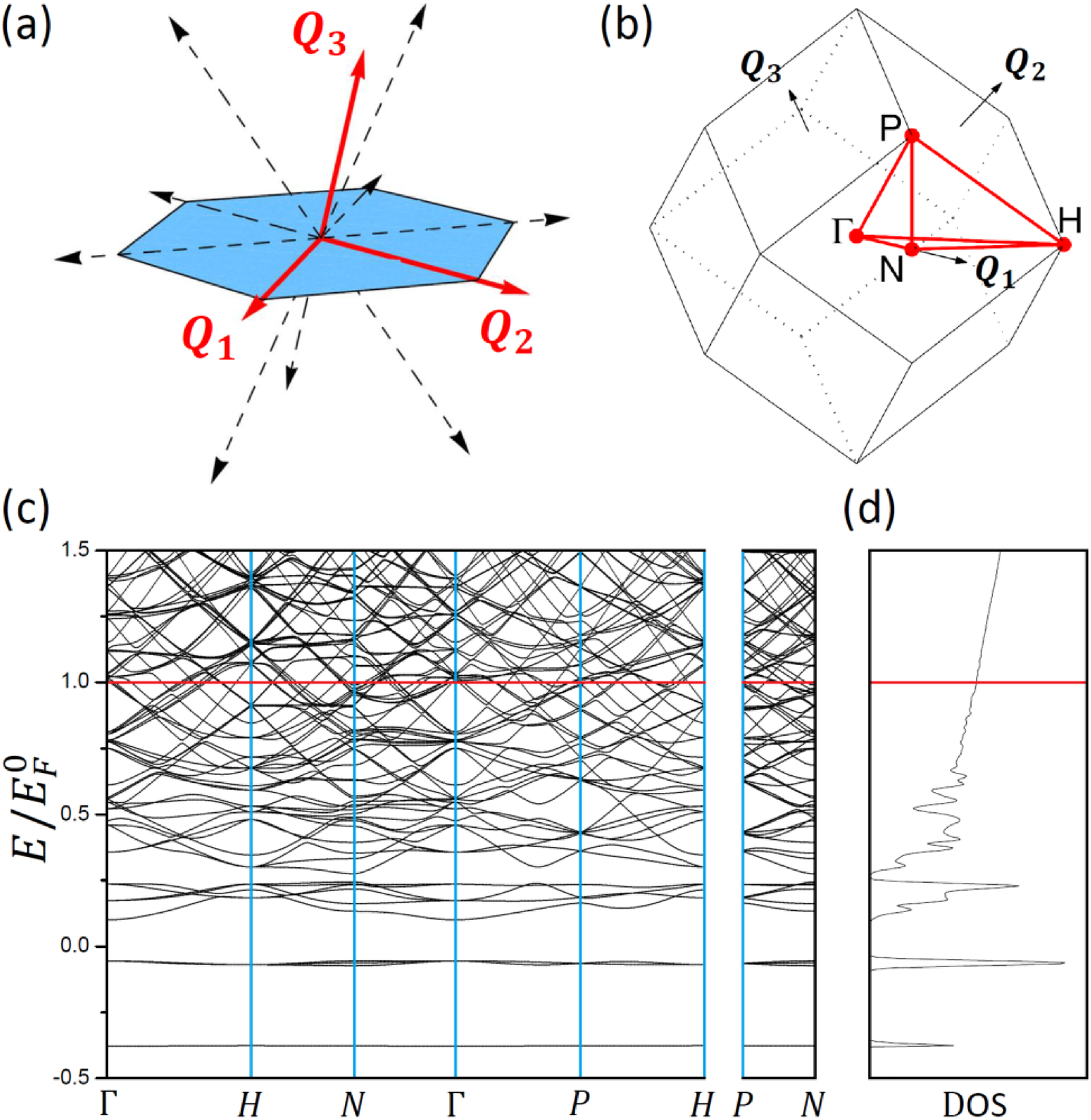} \centering
\caption{(a) Twelve reciprocal vectors of the density wave with a BCC structure, pointing to the 12 reciprocal lattice points closest to the origin. The red solid arrows, $\bfQ_{1,2,3}$, denote the three basis vectors. (b) The first Brillouin zone of BCC lattice with several high symmetry points. (c) The single particle band structure in a metallic density wave state for $R_{c}k_{F}^0=10.62$ and $U_0/E_F^0=0.28$. The horizontal red line indicates the chemical potential of noninteracting system. (d) The corresponding density of states (DOS).}
\label{fig_band_structure}
\end{figure}

{\it Meanfield approach:}
Inspired by the softening of the collective excitation at a finite (but small) momentum, it is reasonable to expect that near the transition boundary a density wave order emerges so that the system becomes non-uniform. As a result, the expectation value of the density operator in momentum space ($\equiv\rho(\bfq)$) can be written to be $\langle\rho(\bfq)\rangle=N\delta _{\mathbf{q},0}+N_{1}\sum_{i=1}^z\delta _{\mathbf{q},\mathbf{Q}_{i}}$, where $N$ is the total particle number and $N_{1}$ is the order parameter of the density wave order. $\bfQ_i$ ($i=1,\cdots,z$) are the wavevectors to describe the density wave order with $|\bfQ_i|=Q_{\rm lat}$. Here we has assumed that the density modulation mostly comes from the first harmonic component for simplicity, and the reciprocal lattice wavevector, $Q_{\rm lat}$, is a variational parameter to be determined by minimizing the total energy. For example, for a cubit lattice, we have $z=6$ and $\bfQ_i/Q_{\rm lat}=\pm \hbfx$, $\pm\hbfy$, $\pm\hbfz$. For a FCC lattice in the momentum space, we have $z=12$ and $\bfQ_1/Q_{\rm lat}=(\hbfx+\hbfy)/\sqrt{2}$, $\bfQ_2/Q_{\rm lat}=(\hbfx+\hbfz)/\sqrt{2}$, $\bfQ_3/Q_{\rm lat}=(\hbfy+\hbfz)/\sqrt{2}$ as the three basis. Other nine $Q_i$'s can be obtained by linear combination of $\bfQ_{1,2,3}$ and has the same amplitude as shown in Fig. \ref{fig_band_structure}(a).

\begin{figure}[tbp]
\includegraphics[width=7cm]{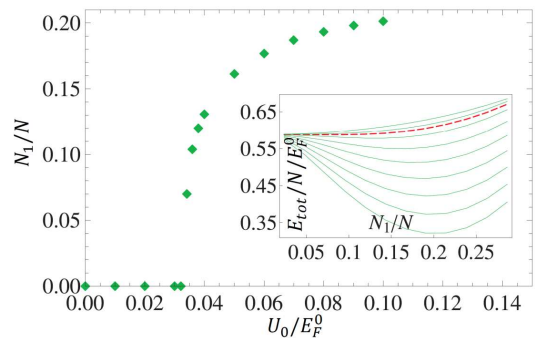} \centering
\caption{
Density wave order parameter, $N_1/N$, as a function of interaction strength, $U_0/E_F^0$, for $k_{F}^0R_{c}\sim 21$. We find $N_{1}/N$ jumps discontinuously at $U_0/E_F^0\sim 0.033$, indicating a first order quantum phase transition. The insert shows the total energy per particle as a function of the order parameter, $N_{1}/N$, for various $U_{0}/E^{0}_{F}$, from $0.03$ (top) to $0.04$ by a step $0.002$, and then $U_{0}/E^{0}_{F}=0.05$, 006, and 0.07 (bottom) respectively. The red dashed line indicates the transition point, which has a local zero energy minimum at $N_1/N\sim 0.07$.}
\label{fig_order_parameter}
\end{figure}

Following the standard meanfield approximation for a charge density wave, we first replace the density operator in the Hamiltonian by its expectation value. Since there is no underlying lattice potential for fermions, we do not expect the nesting effect at any commensurate filling. After neglecting higher order fluctuations, we obtain the following effective meanfield Hamiltonian,
\begin{eqnarray}
H&=&\sum_{\bfk\in 1BZ}\left[\sum_\bfn \varepsilon_{\bfk,\bfn}c_{\bfk,\bfn}^{\dag}c_{\bfk,\bfn}+\frac{N_1V(Q_c)}{\Omega_v}\sum_{\langle \bfn,\bfn' \rangle}
c_{\bfk,\bfn'}^{\dag}c_{\bfk,\bfn}\right]
\nonumber\\
&&-\frac{zN_1^2}{2\Omega_v}V(Q_c),
\label{3dH}
\end{eqnarray}
where we have folded the whole momentum space into the first Brillouin zone ($1BZ$), and defined $c_{\bfk,\bfn}$ to be the field operator at momentum $\bfk$, shifted by a Bravais vector, $\bfK_\bfn\equiv\sum_{i=1}^3n_i\bfQ_i$. Here $\bfQ_{1,2,3}$ are the three primitive vectors in the Bravais lattice, and $\bfn\equiv(n_1,n_2,n_3)$ is an integer vector to label the band index. $\langle\bfn,\bfn'\rangle$ denotes the two neighbouring unit cells, which are coupled by the Rydberg-dressed interaction at a unit Bravais vector, i.e. $|\bfK_\bfn-\bfK_{\bfn'}|=Q_{\rm lat}$. Finally, $\varepsilon_{\bfk,\bfn}\equiv(\bfK_\bfn+\bfk)^2/2m$ denotes the noninteracting band energy, and $\Omega_v$ is the volume of the system.

The meanfield Hamiltonian in Eq(\ref{3dH}) is then diagonalized by a unitary transformation: $\tilde{c}_{\bfk,\bfn}=\sum_{\bfn'}U_{\bfn,\bfn'}^\ast(\bfk) c_{\bfk,\bfn'}$, where $\tlc_{\bfk,\bfn}$ is the eigenstate operator with an eigenenergy, $\tilde{\varepsilon}_{\bfk,\bfn}$. At zero temperature, the total energy for a given chemical potential, $\mu$, can be obtained to be: $E_{\rm tot}(N_1)=\sum_{\bfk,\bfn}\tilde{\varepsilon}_{\bfk,\bfn}\theta(\mu-\tilde{\varepsilon}_{\bfk,\bfn})-\frac{zN_1^2}{2\Omega_v}V(Q_c)$, where $\theta(x)$ is the heaviside function and the chemical potential ($\mu$) is determined by the total particle number. Note that number of atoms in each unit cell of the density wave order needs not to be an interger.  

The order parameter, $N_1$, is then determined self-consistently from the following equation:
\begin{equation}\label{N1}
N_1 =\sum_\bfk \langle   c^\dagger_{\bfk+\bfQ_1}c^{}_\bfk\rangle=\sum_{\bfk,\bfn,\bfm}U_{\bfn,\bfm}U^{\ast}_{\bfm,\bfn+\bfe_x}f(\tilde{\varepsilon}_{\bfk,\bfm}),
\end{equation}
where we set $\bfQ_1=\bfe_x$ and $f(x)=(e^x+1)^{-1}$ is the Fermi distribution function. Note that the transformation matrix element, $U_{\bfn,\bfm}$ depends on $N_1$ also. 

{\it Order parameter and band structure:}
In the numerical calculation, we first consider different types of crystal order and compare their ground energies for different values of $Q_{\rm lat}$. The order parmeter, $N_1$, and chemical potential are calculated self-consistently as described above. After finishing these calculations, we find that a BCC lattice (see the inset of Fig. \ref{collective_mode}(c) with closest sphere packing FCC structure in the reciprocal lattice \cite{Ashcroft}, see Fig. \ref{fig_band_structure}(b)) is energetically most favourable when the density wave order is formed. 

In Fig. \ref{fig_order_parameter}, we show how the order parameter fraction, $N_1/N$, self-consistently calculated from Eq. (\ref{N1}), changes sharply from zero to a finite value as $U_{0}/E_F^0\ge 0.033$ for $R_ck_F^0=21$. In the inset, we show how the total energy changes as a function of $N_1/N$ near the transition point. Both results indicate that the quantum phase transition from a Fermi liquid state to a density wave state is first order, consistent with earlier studies on classical liquid-solid transition \cite{Landau3}. The phase transition boundary is found very close to the results given by roton mode softening (see Fig.\ref{collective_mode}(c)) and the obtained reciprocal lattice wavevector, $Q_{\rm lat}$, is almost the same (within 5\% difference) as the one obtained by roton softening ($Q_c$). Results from these two independent approaches agree very well and hence confirm such new quantum phase transition in a Rydberg-dressed Fermi gas. We further note that since $Q_{\rm lat}\sim Q_c\ll k_F^0$ in the parameter regime to justify our meanfield calculation, It directly implies that there can be many fermionic atoms in each density wave period and hence a new Fermi surface is formed in such lattice structure, leading to a metallic density wave state. 

In Fig.\ref{fig_band_structure}, we further show the the single particle band structures (c) and density of states (d) in the density wave phases for a typical parameter. We note that the elementary excitations near the Fermi surface is essentially gapless, while some band gaps are opened {\it under} the Fermi energy. This results can be understood from the following observation: if only the first band is occupied in the dilute limit (say $R_ck_F^0<3.2$ in our calculation, while the first Born and meanfield approximations may not be justified in this regime), the density wave has an effective one particle per site in the real space. Different from the Wigner in a long-ranged interaction, however, such crystal is incompressible (with a gap) in the single particle excitation, because it costs a finite energy to add one more particle inside the blockage radius, $R_c$, (see the inset of Fig. \ref{Rydberg_energy_level}(b)). For a Coulomb or dipolar interaction, on the other hand, the addition of one more particle can be done without much energy cost by re-arranging the lattice structure to form a defect in the long-wavelength limit, because no intrinsic length scale therein. When adding more particles in each site, Pauli exclusion principle starts to make a Fermi sea and hence a gapless excitation, leading to a metallic quantum solid in the weak interaction regimes we considered here (for the justification of our approximations). 

We can further estimate the critical temperature of such density wave order by solving Eq. (\ref{N1}) with $N_1\to 0$, and find $T_c\sim 0.1E_F^{0}$ near the phase transition boundary. This indicates that the MDW phase proposed above and should be achievable within the present experimental set-up. More sophisticated calculation by including the collective excitations etc. is beyond the scope of this paper, but can be investigated in more details in the future study.

{\it Experimental measurement:}
In a realistic experiment, one can in principle tune the Rydberg-dressed interaction in a very wide range by the external field. Taking $^6Li$ as an example, we find that $C_6\sim 105$ GHz-$\mu$m$^6$ for the 60th orbit, so that  $U_0=9.82$kHz and $R_c=1.27\mu$m by choosing effective Rabi frequency, $\Omega=2\pi\times 100$MHz, and detuning, $\Delta=2\pi\times 2$GHz. For a typical density $n=10^{14}$cm$^{-3}$, we have $U_0/E_F^0=0.036$ and $R_ck_F^0=23.0$, near the phase transition boundary. The obtained metallic density wave has a lattice constant, $2\pi /Q_{\rm lat}=1.51\mu$m, which is about 7 times longer than the average inter-particle distance, $n^{-1/3}\sim 0.21$ $\mu$m.

When considering a Rydberg-dressed Fermi gas trapped in a harmonic potential, one can apply the local density approximation when the cloud size is much larger than the lattice constant, i.e. replacing $k_F^0$ by $k_F(\bfr)=(6\pi^2n(\bfr))^{1/3}$, where the $n(\bfr)$ is the density profile without crystal order. Since the quantum phase transition is first order, we expect the metallic density wave concentrate in the cloud center (higher density) with a discontinuous density change near the interface with the normal surface. These signature should be measurable in the short time of flight experiments.

{\it Conclusion:} We find a new type quantum phase transition of a single component fermionic atoms with a Rydberg-dressed repulsive interaction (say $^6$Li). The observed metallic density wave phase results from the softening of the collective mode excitations with a BCC structure in the 3D real space. We gave an analytic expression of the criteria for such quantum phase and show that the phase transition is of first order. Our results suggest the emergency of a new quantum solid, which should be observable in the future experiment.

We appreciate fruitful discussion with T. Pohl and H.-H. Jen. This work is supported by MoST and NCTS in Taiwan.

\end{document}